\documentclass[useAMS,usenatbib]{mn2e}
 
\usepackage{amsmath}
\usepackage{float}
\usepackage{graphicx}
\usepackage{txfonts}
\usepackage{indentfirst}
\usepackage{color}
\usepackage{ulem}
\usepackage{upgreek}
\let\pi\uppi

\newcommand{\eq}[1]{\begin{equation}  #1 \end{equation}}
\newcommand{\eqs}[1]{\begin{equation} \begin{split} #1 \end{split} \end{equation}}

\newcommand{\items}[1]{\begin{itemize} #1 \end{itemize}}
\newcommand{\br}[1]{\left( #1 \right)}

\newcommand{\bb}[1]{\left[ #1 \right]}

\newcommand{\dd}{{\rm d}}

\def\apj{ApJ}

\def\apjs{ApJS}
\def\aap{A\&A}

\def\mnras{MNRAS}

\def\aplett{Astrophys.~Lett.}

\def\ssr{Space~Sci.~Rev.}

\title[Locations of Accretion Shocks]{Locations of Accretion Shocks around
Galaxy Clusters and the ICM properties: insights from Self-Similar Spherical
Collapse with arbitrary mass accretion rates}

\author[Xun Shi]{Xun Shi \thanks{E-mail:
xun@mpa-garching.mpg.de} \\
Max-Planck-Institut f\"ur Astrophysik,
Karl-Schwarzschild-Stra{\ss}e 1, D-85740 Garching bei M\"unchen, Germany\\
}
 
\begin{document}


\maketitle
  
\label{firstpage}

\begin{abstract}
Accretion shocks around galaxy clusters mark the position where the infalling
diffuse gas is significantly slowed down, heated up, and becomes a part of the
intracluster medium (ICM). They play an important role in setting the ICM
properties.
Hydrodynamical simulations have found an intriguing result that the radial
position of this accretion shock tracks closely the
position of the `splashback radius' of the dark matter, despite the very
different physical processes that gas and dark matter experience. Using the
self-similar spherical collapse model for dark matter and gas, we find that an
alignment between the two radii happens only for a gas with an adiabatic index
of $\gamma \approx 5/3$ and for clusters with moderate mass accretion rates. In
addition, we find that some observed ICM properties, such as the entropy slope
and the effective polytropic index lying around $\sim 1.1-1.2$, are captured by
the self-similar spherical collapse model, and are insensitive to the mass accretion history.

\end{abstract}

\begin{keywords}
galaxies: clusters: general -- galaxies: clusters: intracluster medium --
cosmology: theory -- methods: analytical
\end{keywords}

\section[]{Introduction}
When diffuse gas is accreted on to a galaxy cluster, it inevitably goes through
an accretion shock in the cluster outskirts. 
X-ray and the Sunyaev-Zel'dovich effect observations now start to
reach the sensitivities required to probe the outer regions of galaxy clusters
(e.g. \citealt{planck15XL}, see also \citealt{reiprich13} for a review), and
we may expect a direct detection of the accretion shock in the near future. 
This motivates our studying where we expect to find the accretion
shock, as well as the properties of the intracluster gas in the bulk of the cluster volume.

In hydrodynamical simulations, there exists a complex network of accretion
shocks outside the virial radius of a galaxy cluster due to the
inhomogeneous distribution of matter there \citep{kang07,vazza11,pla13}.
Nevertheless, the peak of the entropy profile provides us with a unique
practical definition of the radial position of the accretion shock
\citep{lau15}. This definition of the shock radius is convenient particularly
because it can also be inferred from observations.
Using this definition, \citet{lau15} found that the shock radius is slightly
larger than, but tracks closely the splashback radius of the dark matter
\citep{diemer14, adhi14, more15} despite of the very different
underlying physics. The two radii align with each other to an extent that
the resulting ratio of gas density and dark matter density stays close to the
cosmic mean value over a large radial range in the cluster outskirts (Fig.\;9 of
\citealt{lau15}), although the density jumps associated with the shock and splashback radii may easily
cause a large deviation.
Moreover, both the accretion shock radius and the splashback radius scale well
with $r_{\rm 200m}$ - the radius within which the averaged density is 200 times
the \textit{mean} density of the universe, rather than the more commonly used
$r_{\rm 200c}$ (for which the reference density is the \textit{critical} density
of the universe) which is better for scaling the cluster inner profiles
\citep{lau15}.

To explore the origin of these curious behaviors, one needs to study the growth
of galaxy clusters in an expanding universe by accreting matter around them. The
self-similar spherical collapse model \citep{bert85} offers us the
simplest analytical framework for doing this. 
With the self-similarity ansatz alone, the self-similar spherical collapse
model allows us to rigorously and consistently derive the self-similarity profiles of the
collapsed regions, e.g. galaxy clusters. This distincts it from the other
smooth accretion models \citep{tozzi01,voit03} which need further
assumptions about the internal structure of the galaxy cluster. 

A somewhat overlooked fact is that the self-similar spherical collapse solutions
given in \citet{bert85} have already captured an alignment between the
accretion shock radius and the splashback radius, as well as some observed ICM
properties such as the slope of the entropy profile and the effective equation
of state of the intracluster gas.
However, \citet{bert85} studied only one particular shape of the initial density
peak which corresponds to a particular mass accretion rate of a galaxy
cluster. Since the mass accretion rate is known to play an important
role in determining the locations of the accretion shock, the splashback radius,
as well as the ICM profiles \citep[e.g.][]{diemer14, adhi14, lau15}, 
it is hard to tell whether the features captured by the \citet{bert85} model
hold at an arbitrary mass accretion rate, or are just coincidences occurring at
that particular mass accretion rate. 

Inspired by the success of the Bertschinger model and motivated by this
shortcoming, we explore the properties of self-similar spherical collapse solutions
systematically by extending Bertschinger's scheme to various mass accretion
rates. We also consider the effect of a different inner mass profile due to
dynamical relaxation, and the dynamical effect of a finite baryon content
(\citealt{bert85} considered only the case of a negligible baryon content, i.e.
$\Omega_{\rm b} \ll \Omega_{\rm m}$).

We will focus on the following physical questions:
\items{
\item  What determines the radial position of the accretion shock? Why does it
track closely the splashback radius of the dark matter?
\item  What sets the entropy slope and the effective equation of state of the
intracluster gas? 
\item The accretion shock slows down the gas compared to the dark matter during
their accretion on to a galaxy cluster - how does this affect the gas mass
fraction inside the cluster? }

We describe the self-similar spherical collapse and its various representations
in Sect.\;\ref{sec:model}. In Sect.\;\ref{sec:rshock} we explore the physical
origin of the accretion shock location, why it aligns well with the dark matter
splashback radius, and how it is affected by a modification of the inner mass
profile and a finite baryon fraction. In Sect.\;\ref{sec:ICMprof} we study the
ICM properties including the entropy profile, the effective equation of state of
the intracluster gas, and the gas mass fraction. We discuss how these results
would be affected by the deviation from smooth
accretion and the existence of dark energy in Sect.\;\ref{sec:discussion}, and
conclude in Sect.\;\ref{sec:con}.

\section[]{Self-similar spherical collapse and its various representations}
\label{sec:model}

We study a secondary infall of a mass shell on to an initial overdense region
that would grow to a halo of the size of a galaxy cluster. The profile of the
initial mass excess we consider is of a power-law shape, $\delta m_i
/ m_i \propto m_i^{-1/s}$. In an Einstein
de-Sitter (EdS) universe, this initial overdensity profile implies a power-law
mass growth $m \propto a^s$ of the halo \citep[e.g.][]{fillmore84}, with $a$ being the cosmic scale factor.
The sizes of these overdense regions are much smaller than that of the
horizon. These, together with a scale-free EdS
cosmological background, ensure that the growth of the halos is self-similar,
as were the cases considered by \citet{fillmore84} and \citet{bert85}.

Self-similarity suggests that the trajectory of a single shell of matter
(matter at a certain Lagrangian radius) describes at the same time the
positions of all the shells in one snapshot, i.e.
the profile of the matter distribution. In fact, the trajectories and profiles
couple in the dynamical equations, and thus need to be solved together. The
solutions can also be presented in both a `trajectory view' and a
`profile view', which complement each other. In the following we present both
views of the self-similar infall process.

\begin{figure*}
\centering
    \includegraphics[width=.97\textwidth]{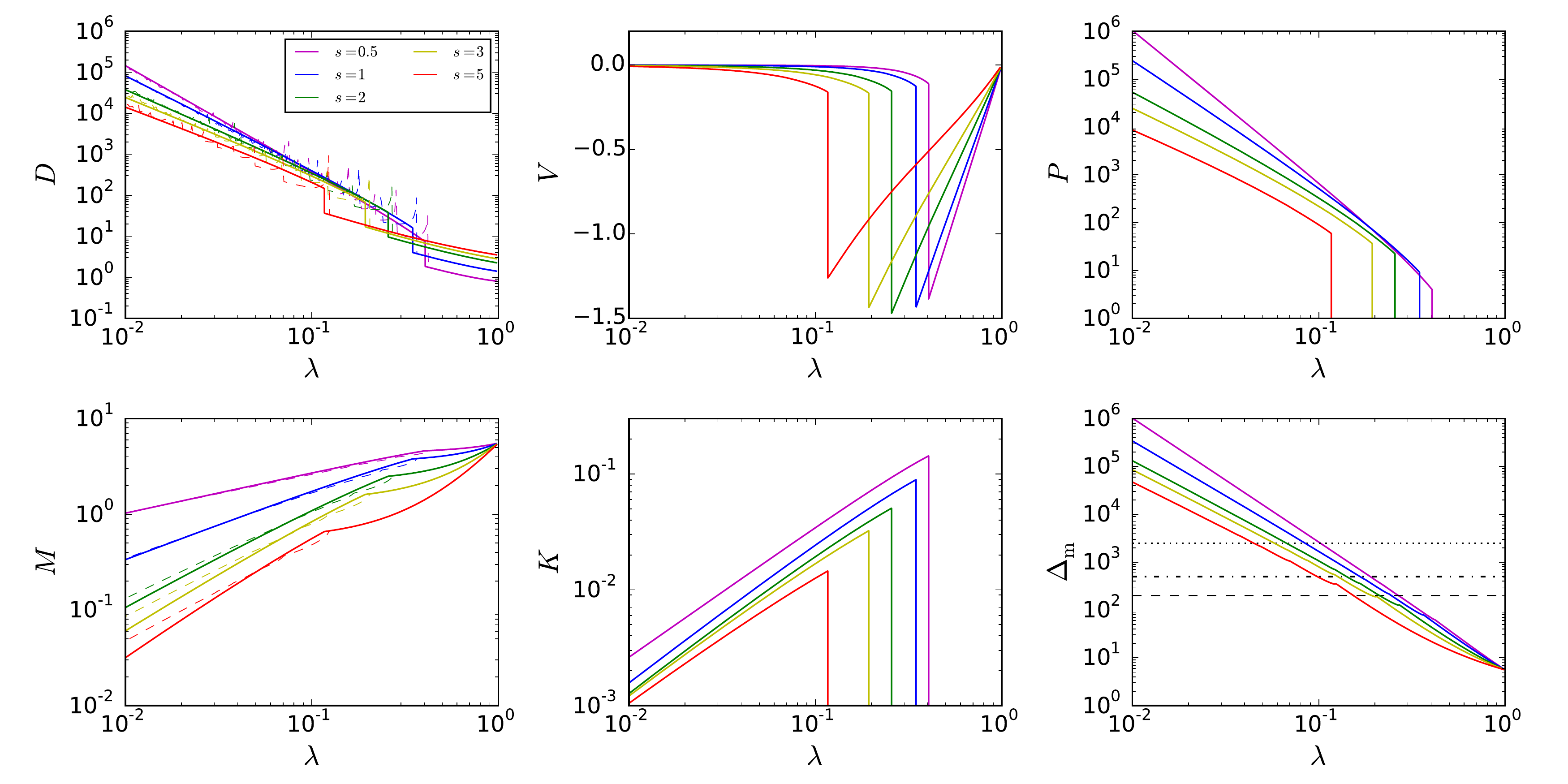} 
  \caption{Radial profiles of the nondimensional gas density (upper left),
  velocity (upper middle), pressure (upper right), mass (lower left), entropy (lower middle), and
  averaged overdensity (lower right) in the self-similar spherical collapse
  model.
  Different colors indicate different mass accretion rates. In the left panels,
  the density and mass profiles of dark matter are plotted as the color dashed
  lines for comparison. The numerical values of the dark matter density
  at the caustics are artificially reduced for the clarity of the general
  profile shape.
  In the lower right panel, averaged overdensity of 200, 500, and 2500 are marked as the dashed,
  dash-dotted, and dotted lines to show the corresponding positions of commonly
  used radii $r_{\rm 200}$, $r_{\rm 500}$ and $r_{\rm 2500}$. The default value
  of $\gamma=5/3$ is taken for the adiabatic index of the gas.}
\label{fig:gasprofs_selfsim}
\end{figure*}

\begin{table*}
 \centering
{\caption{Dimensionless radial position, velocity and acceleration expressed
with different variables.}
\begin{tabular}{ccccc}
 \hline
 variables  & $r$, $t$ &  $\lambda = r/r_{\rm ta}$, $\xi = \ln
(t/t^*)$ & $\lambda_{\rm F} = r/r^*$, $\tau = t/t^*$ &
  $V=vt/r_{\rm ta}$, $\lambda = r/r_{\rm ta}$ \\
\hline
Radial position
&$\frac{1}{r_{\rm ta}(t)} r$
&$\lambda$
&$\tau^{-\delta} \lambda_{\rm F}$
&$\lambda$ \\
Velocity
&$\frac{t}{r_{\rm ta}(t)} \frac{\dd r}{\dd t}$ 	
&$\frac{\dd \lambda}{\dd \xi}  + \delta \lambda$  	
&$\tau^{1-\delta} \frac{\dd \lambda_{\rm F}}{\dd \tau}$ 	
&$V$ \\
Acceleration
&$\frac{t^2}{r_{\rm ta}(t)} \frac{\dd^2 r}{\dd t^2}$ 	
&$\frac{\dd^2 \lambda}{\dd \xi^2} + \br{2\delta-1} \frac{\dd \lambda}{\dd \xi} +
\delta \br{\delta-1} \lambda$  	
&$\tau^{2-\delta} \frac{\dd^2 \lambda_{\rm F}}{\dd \tau^2}$ 	
&$(V-\delta\lambda)V' + (\delta-1)V$ \\	
\hline
\end{tabular}}
\label{tab:tabvforms}
\end{table*}

We consider the self-similar infall trajectories of cold dark matter and gas.
A spherical shell of gas obeys the equation of motion 
\eq{
\label{eq:eom}
\frac{\dd^2 r}{\dd t^2} = -\frac{G m}{r^2} - \frac{1}{\rho}\frac{\partial
p}{\partial r} \,,
}
while for dark matter there is no pressure
force (the second term on the r.h.s.).
The mass $m$ represents the total mass of dark matter and gas. We consider an
initial gas mass fraction equal to a cosmic mean value $\Omega_{\rm
b}/\Omega_{\rm m}$. At first we consider $\Omega_{\rm
b}=0$, i.e. assume that the mass is fully dominated by the dark matter, and
study the effect of a finite baryon content $\Omega_{\rm
b}$ in Sect.\;\ref{sec:diffprof}.

As the typical temperature of the intergalactic diffuse gas is several
orders of magnitude lower than that of the intracluster gas, we neglect the
temperature and pressure in the gas before it gets shocked upon entering
the pre-existing halo. With this cold-accretion assumption, gas and dark matter
are dynamically indistinguishable before the accretion shock. The
enclosed total mass stays at its initial value $m_i$ during their initial
expansion with the universe until the common trajectory turns around at a time
$t^*$ and radius $r^*$ due to the excess mass contained in $m_i$.
After the turn-around, the gas experiences an accretion shock at
some radius, which slows down and separates it dynamically from the
collisionless dark matter.

\citet{bert85} realised that, thanks to self-similarity, the total mass profile
must have a universal shape when normalised to the values at the current
turn-around radius $r_{\rm ta}$\footnote{We distinguish the quantities at the turn-around radius
\textit{of the shell} which we indicate using superscript `*', to the quantities
at the `current turn-around radius' (the turn-around radius \textit{at the time
of observation}, i.e. $t^*=t^{\rm obs}$) which we indicate with subscript
`ta'.}.
The same holds true for the individual gas and dark matter mass and density
profiles as well as for the velocity and thermodynamical profiles of the gas.
We define the nondimensional profiles of gas velocity $V$, density $D$,
pressure $P$, and the total mass profile $M$ as 
\eqs{
\label{eq:nondim}
v & = \frac{r_{\rm ta}}{t} V(\lambda) \,,\\
\rho & = \rho_{\rm H} D(\lambda) \,,\\
p & = \rho_{\rm H} \br{\frac{r_{\rm ta}}{t}}^2 P(\lambda) \,,\\
m & = \frac{4\pi \rho_{\rm H}}{3} r_{\rm ta}^3 M(\lambda)  \,,
}
with $\lambda = r / r_{\rm ta}$ and $\rho_{\rm H}$ being the mean matter density
in the universe, $\rho_{\rm H}  = 1/(6\pi G t^2)$. The equation of motion
Eq.\;(\ref{eq:eom}) can then be written in a form that has no explicit time
dependence \eq{ \frac{\dd^2 \lambda}{\dd \xi^2} + \br{2\delta-1} \frac{\dd
\lambda}{\dd \xi} +
\delta \br{\delta-1} \lambda = 
-\frac{2}{9} \frac{M(\lambda)}{\lambda^2}  -\frac{P'}{D} \,,
\label{eq:eom2}
}
where the prime denotes taking derivative with respect to $\lambda$, and $\xi =
\ln (t/t^*)$ is the new time variable.
We have extended Bertschinger's original equation to an arbitrary mass growth
rate $s$.
Here, $\delta = 2(1 + s/3)/3$ bears the meaning of the time-dependence of the current turn-around radius, i.e. $r_{\rm ta} \propto
t^{\delta}$. This time-dependence of $r_{\rm ta}$ gives rise to the additional
$\dd \lambda / \dd \xi$ and $\lambda$ terms on the l.h.s. of
Eq.\;(\ref{eq:eom2}). A natural choice of the initial condition for
Eq.\;(\ref{eq:eom2}) is $\lambda=1$ and $\dd \lambda / \dd \xi = \delta$ at turn-around, i.e. at $\xi=0$. To help visualising
how different forms of the dynamical equations match with each other, we
summarise the expressions for the radial position, velocity and acceleration in
terms of different variables in Table.\;1.

Since dark matter is collisionless, a dark matter shell falls all the way to the
center of the cluster and then continues to move outward until reaching the
so-called splashback radius $\lambda_{\rm sp}$, and turns around again. The dark matter mass at
$\lambda < \lambda_{\rm sp}$ is then contributed by multiple dark matter
streams accreted at different times.

Deriving the dark matter mass profile requires solving the pressure-less form
of the equation of motion Eq.\;(\ref{eq:eom2}) iteratively together with    
\eq{
\label{eq:mdm}
M(\lambda) 
= M_{\rm ta} \sum_{i} (-1)^{i-1} \rm{exp}
\bb{-(2s/3) \xi_i}  \,,
}
where $\xi_i$ is the value of $\xi$ at the $i$th point with
$\lambda=\lambda(\xi)$ \citep{bert83,bert85}. The alternating signs are related
to adding/subtracting the contribution from dark matter streams moving inward/outward at radius $\lambda$. The $2s/3$ factor in the exponent is related to
the fact that the mass enclosed in a certain radius $\lambda$ grows
with time as $m(\lambda r_{\rm ta}, t) \propto t^{2s/3}$. The $M_{\rm ta} =
\br{{3\pi}/{4}}^2 \approx 5.55$ prefactor is the overdensity at turn-around
\citep[e.g.][]{lacey93}.

As for gas, we seek the solution of a
single self-similar accretion shock that stays at a fixed $\lambda_{\rm sh}$
\citep{bert85}.
In the Eulerian coordinate this corresponds to an outward propagating shock front with
velocity $v_{\rm sh} = \delta \lambda_{\rm sh} r_{\rm ta} / t$. For
$\lambda>\lambda_{\rm sh}$, i.e. before the shock passage, the gas velocity is
derivable from the trajectory as $V = \dd \lambda / \dd \xi + \delta \lambda$,
and its mass is $M_{\rm gas}(\lambda) = M_{\rm ta} \rm{exp} \bb{-(2s/3)
\xi_1}$. 

After going through the accretion shock, the gas obtains a non-negligible
pressure whose dynamical effect needs to be accounted for in the fluid equations
\eqs{
&\frac{\dd \rho}{\dd t} = -\frac{\rho}{r^2}\frac{\partial}{\partial r} \br{r^2v}
\,,
\\
&\frac{\dd v}{\dd t} =  - \frac{Gm}{r^2} -\frac{1}{\rho}\frac{\partial
p}{\partial r} \,, \\
& \frac{\dd}{\dd t}\br{p \rho^{-\gamma}} = 0  \,,\\
& \frac{\partial m_{\rm gas}}{\partial r} = 4\pi r^2 \rho \,.
}
We have assumed that the gas is adiabatic with an equation of
state $\gamma$.
As a default we consider $\gamma=5/3$, which is the value for mono-atomic ideal
gas. The nondimensional forms of these continuity, Euler, adiabatic and mass equations are 
\eqs{
\label{eq:fluid}
& \bb{V-\delta\lambda} D' + DV' + \frac{2DV}{\lambda} -2D
= 0 \,,\\ 
& \bb{V-\delta\lambda} V' + (\delta-1)V = 
- \frac{2}{9}\frac{M}{\lambda^2} -\frac{P'}{D} \,,\\
& \bb{V-\delta\lambda} \br{\frac{P'}{P} -
\gamma\frac{D'}{D}} = -2(\gamma-1) - 2(\delta-1) \,,\\
& M'_{\rm gas} = 3\lambda^2 D \,.
}
They are solved with the outer boundary conditions at the accretion shock given
by the shock jump conditions 
\eqs{
\label{eq:shock}
V_2 &=  \frac{\gamma-1}{\gamma+1}
\bb{V_1-\delta\lambda_{\rm sh}} +
\delta\lambda_{\rm sh} \,,\\
D_2 &= \frac{\gamma+1}{\gamma-1} D_1 \,,\\
P_2 &= \frac{2}{\gamma+1} D_1 \bb{V_1-\delta\lambda_{\rm
sh}}^2  \,,\\
M_2 &= M_1 \,.
}
The pre-shock velocity $V_1$ and density $D_1$ are given by the pre-shock gas
velocity and mass profiles once the shock radius $\lambda_{\rm sh}$ is known.
The shock radius is in turn determined by requiring
the solutions of Eq.\;(\ref{eq:fluid}) to satisfy the inner boundary conditions
$V(0) = 0$ and $M_{\rm gas}(0) = 0$. 

The resulting gas profiles are shown in Fig.\;\ref{fig:gasprofs_selfsim} for
various mass accretion rates $s=0.5,1,2,3$ and $5$. The nondimensional
profiles of the gas density (upper left), velocity (upper middle), pressure
(upper right), mass (lower left), entropy (lower middle), and averaged overdensity (lower right)  
all depend on the accretion rate $s$. In particular, the radial position of the
accretion shock significantly decreases with an increasing accretion rate. The
shapes of the profiles, on the other hand, show some regularity not sensitive to
the accretion rate. For instance, the slope of the entropy profile remains
rather unchanged with $s$, and the slopes of the density, pressure and mass
profiles within the accretion shock saturate to a certain value at high
accretion rates. We will discuss these behaviors in more detail in
Sects.\;\ref{sec:rshock} and \ref{sec:ICMprof}.

\begin{figure}
\centering
    \includegraphics[width=.47\textwidth]{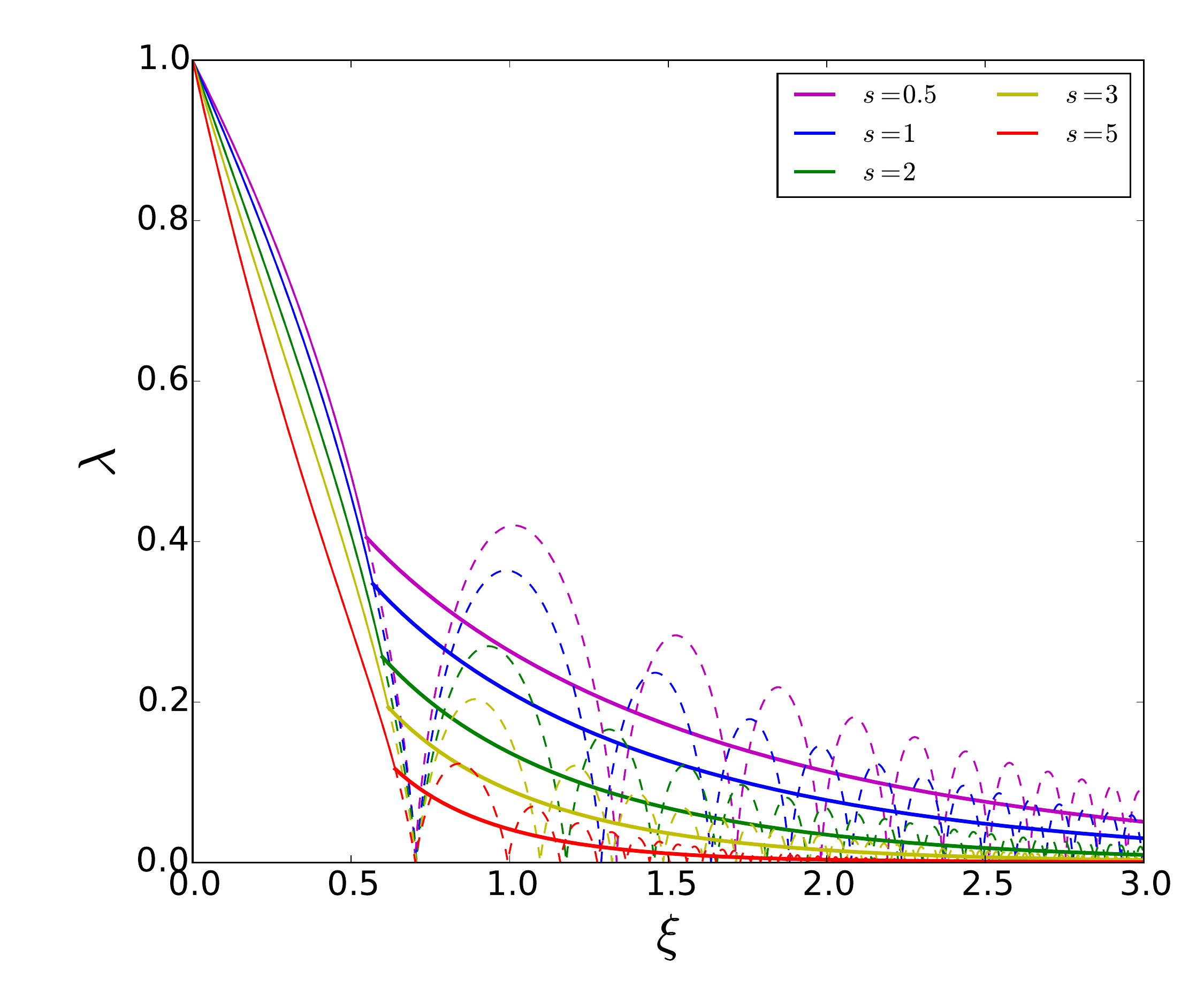} 
  \caption{Trajectories of self-similar spherical infall of dark matter (dashed
  lines) and gas (solid lines) starting from turn-around. Infall trajectories
  with various mass accretion rates $s$ are shown as the lines of different
  colors.
  The radii of the accreted shells are normalized to the current turn-around
  radius, i.e. $\lambda = r/r_{\rm ta}$, and are plotted against the logarithmic
  time $\xi=\ln (t/t^*)$.}
\label{fig:traj1}
\end{figure}

\begin{figure}
\centering
  \begin{tabular}{@{}c}
    \includegraphics[width=0.47\textwidth]{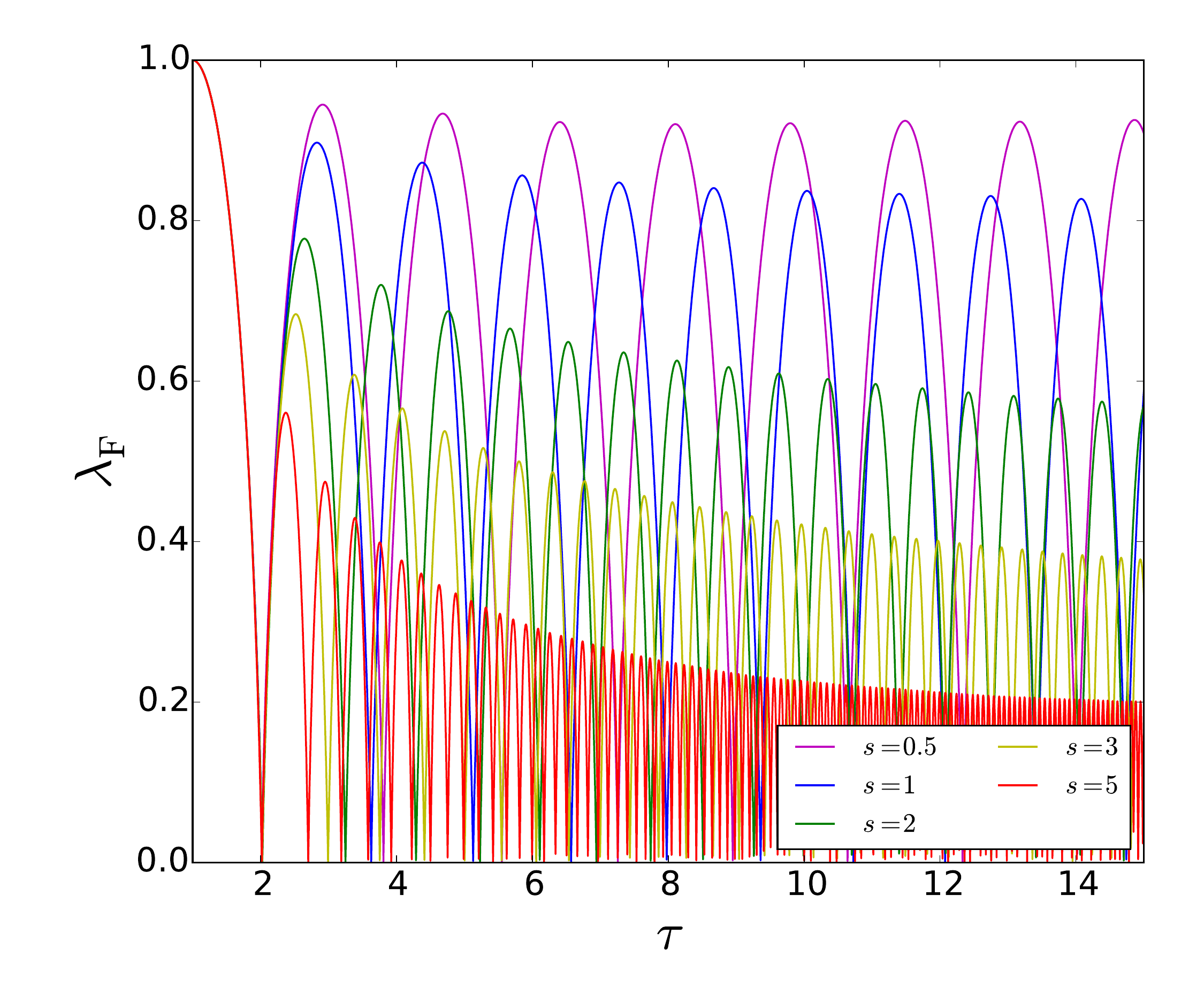}\\   
    \includegraphics[width=0.47\textwidth]{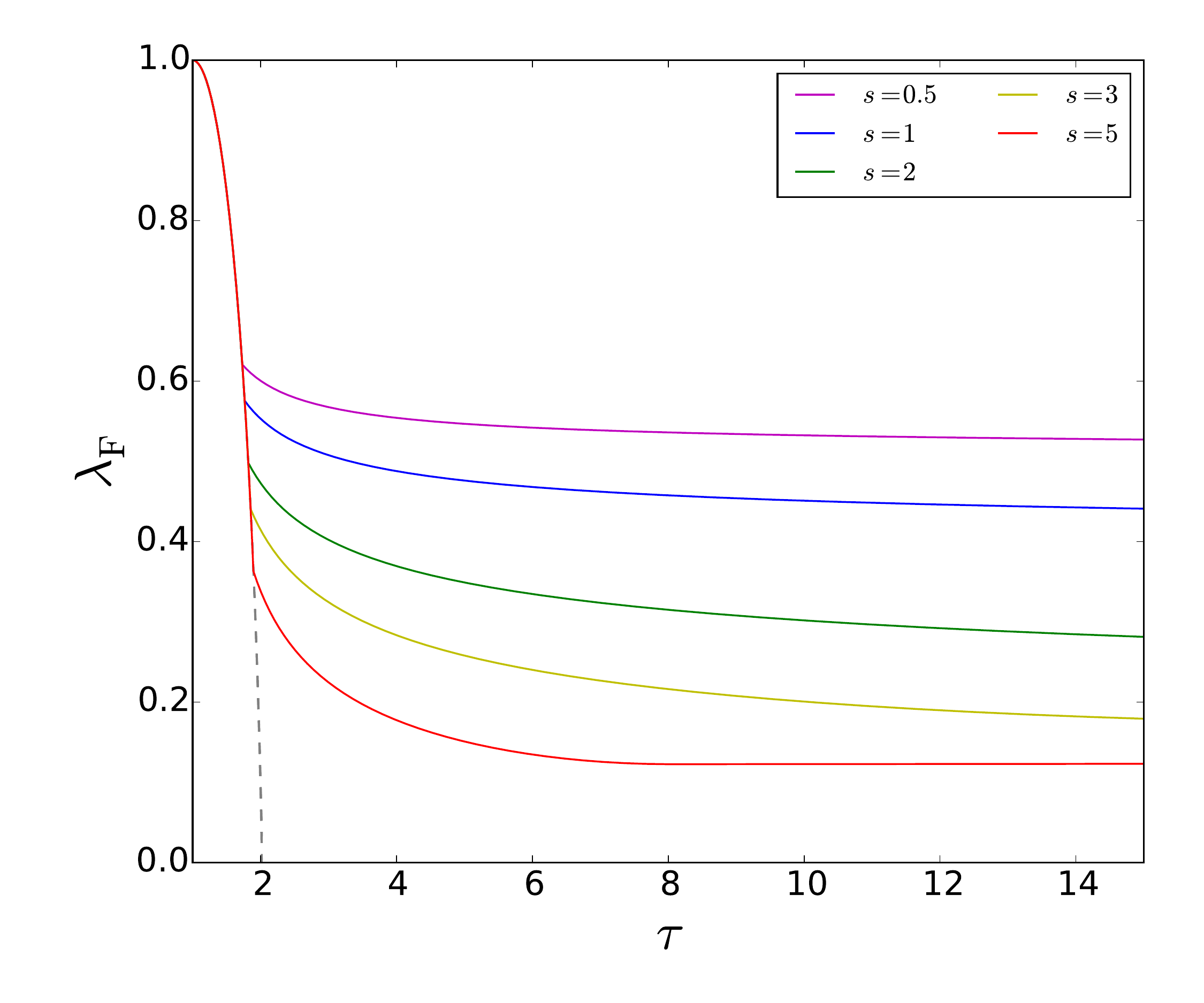}
  \end{tabular}
\caption{Another view of the trajectories of self-similar infall of dark matter
(upper panel) and gas (lower panel). Same as Fig.\;\ref{fig:traj1} but with the radii of the
accreted shells normalized to their turn-around radius, i.e. $\lambda_{\rm F} =
r/r^*$, and the time normalized to their turn-around time, i.e. $\tau = t/t^*$.
The gray dashed line in the lower panel shows the trajectory for the dark
matter for a comparison.}
\label{fig:traj2}
\end{figure}

We also present the `trajectory view' of the self-similarity solutions in
Figs.\;\ref{fig:traj1} and \ref{fig:traj2}, where the trajectories of dark
matter and gas are normalised to the current turn-around radius $r_{\rm ta}$ and
the turn-around radius of the shell $r^*$, respectively. From
Fig.\;\ref{fig:traj1} one can clearly read out the splashback radius for the
dark matter (the first peaks of the dashed lines after turn-around) and the
accretion shock radius for the gas (where the solid lines start to deviate from
the dashed lines). Both the splashback radius and the shock radius decrease
from the turn-around radius with an increasing mass accretion rate $s$.
These two radii follow each other closely for any accretion rate (indicated by the color of the lines). It
is remarkable that this behavior, which was
discovered previously in adiabatic hydrodynamical simulations for galaxy
clusters \citep{lau15}, is already captured by the self-similarity
solutions. This allows us to explore the origin of this alignment using a simple
analytical framework, which we will do in Sect.\;\ref{sec:rshock}.

When normalized to the current turn-around radius, the trajectories of both dark
matter and gas reduce in amplitude with time (Fig.\;\ref{fig:traj1}) due to the
expansion of the current turn-around radius as the universe expands. In the
Eulerian coordinate, however, the amplitudes of these trajectories do not
necessarily shrink with time. Fig.\;\ref{fig:traj2} presents the trajectories
normalized to their own turn-around radius $r^*$ and time $t^*$ by plotting
$\lambda_{\rm F} = r/r^*$ against $\tau = t/t^* = \exp(\xi)$ \citep{fillmore84}. 
Over a long time scale, the amplitudes of the oscillating
trajectories for dark matter continue to shrink due to the increase of mass within an Eulerian radius
when $s>3/2$ \citep{fillmore84}, while remains constant when $s\le 3/2$ (upper
panel).
The gas, on the other hand, always settles at a finite Eulerian radius within its
shock radius (lower panel).

\section[]{Radial position of the accretion shock}
\label{sec:rshock}

\subsection[]{Physical origin}
\label{sec:origin_rsh}

\begin{figure}
\centering
    \includegraphics[width=.47\textwidth]{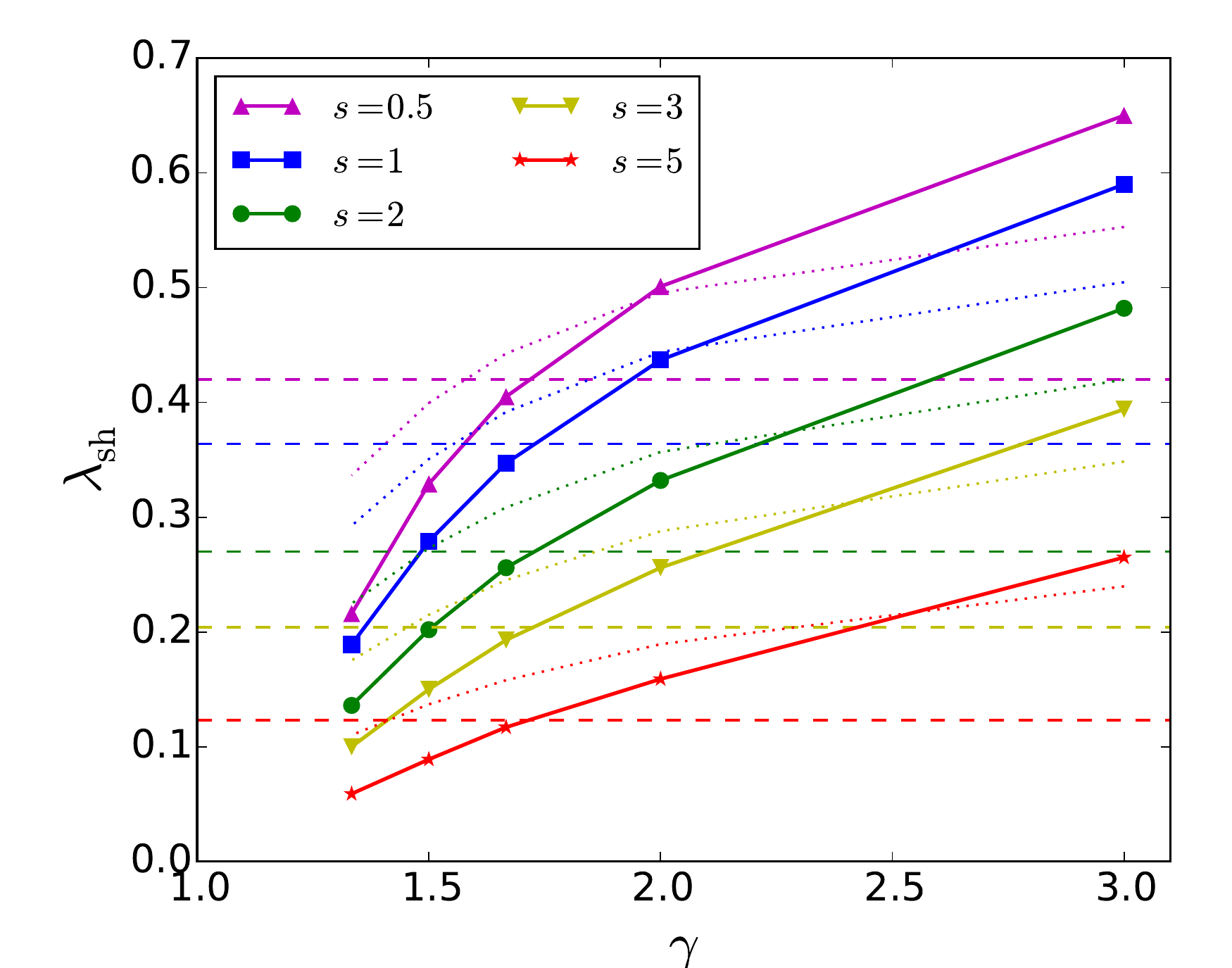} 
  \caption{Positions of the accretion shock for gas with various adiabatic
  indices $\gamma$, and for various accretion rates $s$ indicated by the color.
  The splashback positions of the collisionless matter are marked by the dashed
  lines. The dotted lines show the approximate estimation of the shock position using Eqs.\;(\ref{eq:Econserv_infall}) and (\ref{eq:v2approx0}).}
\label{fig:shockpos_gamma}
\end{figure}

\begin{table*}
 \centering
{
\caption{Tabular form of Fig.\;\ref{fig:shockpos_gamma}. Positions of
the accretion shock for gas, $\lambda_{\rm sh} = r_{\rm sh} / r_{\rm ta}$, with
different equations of state $\gamma$ (columns 2-6) and the splashback position,
$\lambda_{\rm sp}$, of the collisionless matter (column 7) for various
accretion rates $s$.
}
\begin{tabular}{ccccccc}
 \hline
  & $\gamma=3$ & $\gamma=2$ & $\gamma=5/3$  & $\gamma=3/2$ & $\gamma=4/3$ & Collisionless, splashback\\
\hline
$s=0.5$  &$0.650$ &$0.501$  & $0.405$	&$0.329$	&$0.216$ 	&$0.420$ \\
$s=1$  	 &$0.590$ &$0.437$  & $0.347$	&$0.279$ 	&$0.189$ 	&$0.364$ \\ 
$s=2$ 	 &$0.482$ &$0.332$  & $0.256$	&$0.202$	&$0.136$ 	&$0.270$ \\
$s=3$	 &$0.394$ &$0.256$  & $0.193$	&$0.150$ 	&$0.100$ 	&$0.204$ \\	
$s=5$	 &$0.265$ &$0.159$ 	& $0.117$	&$0.089$ 	&$0.059$ 	&$0.123$ \\	
\hline
\end{tabular}}
\label{tab:lambd_sh_gamma}
\end{table*}

We show the dependencies of the accretion shock position on the adiabatic index
$\gamma$ and the accretion rate $s$ in Fig.\;\ref{fig:shockpos_gamma}. The
accretion shock position $\lambda_{\rm sh}$ is larger for a higher accretion
rate, and for a more stiff gas with a higher $\gamma$ value.

What determines the radial position of the accretion shock? 
Strictly speaking, the radial position of the accretion shock $\lambda_{\rm
sh}$ in the self-similar solutions is derived as an eigenvalue of the
fluid equations by requiring them to satisfy the inner boundary conditions at $r=0$. 
To gain physical insights, however, it is better to take
the point of view of the infalling gas, in which the position where it will get
shocked is specified by the shock jump conditions Eq.\;(\ref{eq:shock}). 

The shock jump conditions express the continuity of mass, momentum and enthalpy
across the shock front. Given the properties of shock-bounded matter inside
the halo, the accretion shock must occur at a position where the
properties of the inflowing matter allow these continuities. Here, we use the
simple framework of self-similar spherical collapse model to give
approximations for both the pre-shock flow and the post-shock halo gas, and show
that combining the two indeed sets the location of the accretion shock. 

The infall velocity of the pre-shock flow relates to the shock position
$\lambda_{\rm sh}$ through the conservation of the sum of 
kinetic and gravitational energy of the infalling gas. Namely, the kinetic
energy at the position of the accretion shock should equal the difference of
the gravitational energies at the accretion shock and the turn-around locations.
Expressed in a dimensionless form, it is 
\eq{
\label{eq:Econserv_infall_F}
\frac{1}{2}\br{\frac{\dd \lambda_{\rm F}}{\dd \tau}}_{\rm sh}^2 = \frac{2 M_{\rm
ta}}{9} \br{\frac{1}{\lambda_{\rm F, sh}} -1} \,,
}
or in terms of the dimensionless pre-shock velocity $V_1$ (see Table\;1), 
\eq{
\label{eq:Econserv_infall}
\frac{1}{2} V_1^2 \tau_{\rm sh}^{2\delta-2} = \frac{2 M_{\rm ta}}{9}
\br{\frac{1}{\lambda_{\rm sh}\tau_{\rm sh}^{\delta}} -1} \,.
}
Here we have taken the mass enclosed in the infalling shell
to be a constant, which holds unless accretion shock
happens significantly inside the splashback radius of the dark matter.

The properties of the post-shock halo gas are harder to approximate to a good
precision with a simple analytical form. For a very crude estimation,
we take the approximation $V_2 \approx 0$ which would mean that the post-shock gas is
at perfect hydrostatic equilibrium. This approximation suggests (from the
first equation in the shock jump conditions Eq.\;\ref{eq:shock}) 
\eq{
\label{eq:v2approx0}
\br{V_1 - \delta \lambda_{\rm sh}} \frac{\gamma-1}{\gamma+1} + \delta
\lambda_{\rm sh} \approx 0 \,. 
}
In the actual solutions, some residue gas motion still exists after the
accretion shock (see the upper middle panel of
Fig.\;\ref{fig:gasprofs_selfsim}). 

Now, combining Eqs.\;(\ref{eq:Econserv_infall}) and (\ref{eq:v2approx0}) allows
us to derive an approximation for the accretion shock location $\lambda_{\rm sh}$.
Despite the approximations, the result gives the correct trends of how
$\lambda_{\rm sh}$ depends on the control parameters $\delta$ and $\gamma$ (see
the dotted lines in Fig.\;\ref{fig:shockpos_gamma})\footnote{For the estimation, we fix $\tau_{\rm sh}=2$ which is a rough approximation based on the symmetry of expansion and infall around the turn-around point (by definition, $\tau=1$ at turn-around). The precise value of $\tau_{\rm sh}$ is
slightly smaller than 2 and varies with the accretion rate and the
adiabatic index of the gas (see Fig.\;\ref{fig:traj2}) }, confirming that the
accretion shock position is determined from the continuities of the halo
gas which is approximately in hydrostatic equilibrium and the pre-shock gas
which carries kinetic energy gained during the gravitational infall. It
also helps us to understand the parameter dependencies
(Fig.\;\ref{fig:shockpos_gamma} and Table.\;2):
both a smaller adiabatic index $\gamma$ and a larger accretion rate $s$ require a larger pre-shock velocity
$V_1$ at a certain shock position $\lambda_{\rm sh}$ to satisfy the shock
jump conditions Eq.\;(\ref{eq:v2approx0}), which implies a smaller $\lambda_{\rm
sh}$ since it allows more conversion of the gravitational energy
(Eq.\;\ref{eq:Econserv_infall}) and thus gives a higher pre-shock velocity.

\subsection[]{Comparison with the dark matter splashback radius}
\label{sec:2rs}

State-of-the-art hydrodynamical numerical simulations have found the curious
result that the accretion shock radius of the gas and the splashback radius of the
dark matter follow each other closely \citep{lau15} despite the very
different physical processes involved. Is this a pure coincidence or is there
some physical reason behind? 

It was already noticed by \citet{bert85} that this result holds true only for a
gas with an adiabatic index $\gamma \approx 5/3$, for an accretion rate he
studied. We extend his analysis to an arbitrary value of the accretion rate.
In Fig.\;\ref{fig:shockpos_gamma}, we show the positions of the splashback
radius with the horizontal dashed lines. We find that, for and only for $\gamma
\approx 5/3$, the accretion shock position aligns well with the splashback
radius for all the tested accretion rates. The mystery is then two-folds:
why for $\gamma \approx 5/3$, and more intriguingly, why does this alignment
hold at $\gamma \approx 5/3$ for any accretion rate?

To find the answers we now consider the origin of the splashback radius of the
dark matter. Initially, the dark matter shares the same expansion and infall
with the gas and obeys Eq.\;(\ref{eq:Econserv_infall_F}). It then enters the
multi-stream region when it reaches $\lambda_{\rm sp}$, and starts to oscillate
around the cluster center in the potential well of the cluster. The
approximate position of the splashback radius can be given analytically by
considering this process, yielding \citep{shi16}
\eq{
 \label{eq:lambda_sp}
  \lambda_{\rm sp} \approx
  \begin{cases}
    3^{-\delta}  & \quad \text{if } \delta \le 1 \ \,(s \le 3/2) \,,\\
    \bb{1 + 4(2\delta-1)/\sqrt{\pi}}^{-1}       & \quad \text{if } \delta \ge
    1\ \, (s \ge 3/2) \,.\\
  \end{cases}
}
Once again, $\delta = 2(1+s/3)/3$.
The two regimes of accretion $\delta \le 1$ and $\delta > 1$ correspond to an
approximately constant inner gravitational potential well and one growing with
time, respectively. 
Due to the increase of the turn-around radius $r_{\rm ta}$ with time, the
amplitude of the oscillation in terms of $\lambda$ decreases with time
(Fig.\;\ref{fig:traj1}) even when $\delta \le 1$, and thus $\lambda_{\rm sp}<1$
for all non-vanishing mass accretion rates.

The alignment of the two radii holds for an adiabatic gas with $\gamma
\approx 5/3$ for $0.5 \le s \le 5$ suggests that for this range of mass
accretion rates, the dynamical effect of the accretion shock on a $\gamma
\approx 5/3$ gas mimicks that of the growth of the cluster and $r_{\rm ta}$ on the dark
matter. The value $\gamma \approx 5/3$ reflects nothing more fundamental, and
may not apply for more extreme mass accretion rates either. In fact, combining
the approximate expressions for $\lambda_{\rm sh}$ (Sect.\;\ref{sec:origin_rsh}) and $\lambda_{\rm sh}$ (Eq.\;\ref{eq:lambda_sp})
suggests that for higher mass accretion rates $s>6$, harder adiabatic
indices ($\gamma > 5/3$) are required for the accretion shock position
to align with the splashback radius.

Both the accretion shock and the splashback radius decrease with a higher mass
accretion rate but for different reasons.
For the gas, it is because a higher mass accretion rate $s$ suggests higher energy and momentum of
the inflowing gas, whereas for the dark matter, it is due to a more significant
cluster growth during the time between the splashback and when the matter at
splashback was accreted. 
That their dependencies on the mass accretion rate are so similar
for the range of accretion rates we tested is likely just a coincidence.

\subsection[]{Effect of relaxation and $\Omega_{\rm b}$}
\label{sec:diffprof}

\begin{figure}
\centering
    \includegraphics[width=.42\textwidth]{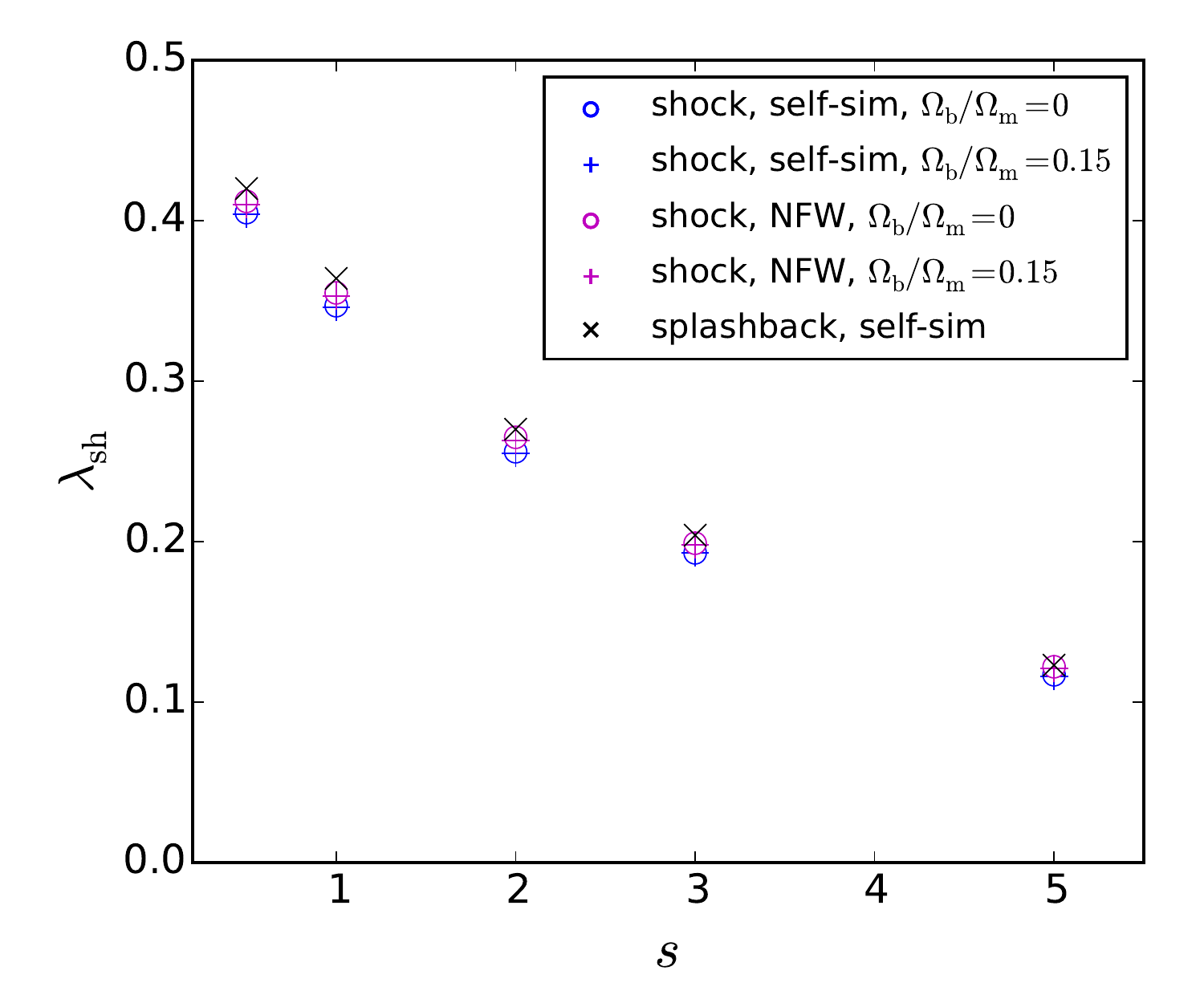} 
  \caption{Positions of the accretion shock of an adiabatic gas with
  $\gamma=5/3$ compared to the splashback position of dark matter (black crosses) for various accretion rates
  $s$. The shock radius is computed with various 
  total mass profiles,
  including a self-similar profile (blue markers) and that with inner profile modified to an NFW shape
  (magenta markers), both with $\Omega_{\rm b}/\Omega_{\rm m}=0$ (`o') or
  $\Omega_{\rm b}/\Omega_{\rm m}=0.15$ (`+').
  }
\label{fig:lambd_sh}
\end{figure}

\begin{table*}
 \centering
{
\caption{Tabular form of Fig.\;\ref{fig:lambd_sh}. Positions of the accretion
shock (columns 2-5), $\lambda_{\rm sh}$, compared to the splashback position of
the collisionless matter (columns 6), $\lambda_{\rm sp}$, for various accretion
rate $s$.
}
\begin{tabular}{cccccc}
 \hline
  & Self-similar, $\Omega_{\rm b}/\Omega_{\rm m}=0$  & Self-similar, $\Omega_{\rm b}/\Omega_{\rm m}=0.15$ & NFW,
  $\Omega_{\rm b}/\Omega_{\rm m}=0$ & NFW,  $\Omega_{\rm b}/\Omega_{\rm m}=0.15$
  & Self-similar, splashback\\
\hline
$s=0.5$   		& $0.405$	&$0.404$	&$0.412$ 	&$0.410$ 	&$0.420$ \\
$s=1$  			& $0.347$	&$0.346$ 	&$0.355$  	&$0.353$ 	&$0.364$ \\ 
$s=2$ 			& $0.256$	&$0.255$	&$0.265$ 	&$0.263$ 	&$0.270$ \\
$s=3$			& $0.193$	&$0.193$ 	&$0.199$  	&$0.198$ 	&$0.204$ \\	
$s=5$			& $0.117$	&$0.116$ 	&$0.122$  	&$0.121$ 	&$0.123$ \\	
\hline
\end{tabular}}
\label{tab:lambd_sh_sp}
\end{table*}

How representative is the accretion shock radius from the $\Omega_{\rm b}
= 0$ self-similar solution that has a power-law inner mass profile? Here we
explore the influence of a finite baryon content $\Omega_{\rm b}/\Omega_{\rm m}=0.15$ close to the observed values
\citep{kom11,planck15XIII} and a more realistic NFW mass profile
\citep{nfw96,nfw97} on the accretion shock position.
Both the two modifications should influence the accretion shock position through
their change to the total mass profile. Since the self-similar
solutions of the gas and dark matter mass profiles in the $\Omega_{\rm b} = 0$
case are very similar for a $\gamma=5/3$ gas (Fig.\;\ref{fig:gasprofs_selfsim}),
the change to the total mass profile by a moderate baryon content is expected to be
small, and thus a direct modification of the shape of the inner mass profile to
an NFW profile is expected to give a greater change to the accretion shock position. 

The way we embed an NFW profile in the inner region of the halo follows that
of \citet{adhi14}. We determine the concentration
parameter $c$ by requesting that the slope of the NFW profile at the virial
radius is equal to the inner mass slope $\Upsilon$ of the self-similar
solution, i.e. $c^2/(1+c)^2/m_{\rm nfw}(c) = \Upsilon$ with 
$m_{\rm nfw}(x) = \ln(1+x) - x/(1+x)$, and the asymptotic inner dark matter mass
slope as derived by \citet{fillmore84} 
\eq{
 \label{eq:slope_dmmass}
  \Upsilon =
  \begin{cases}
    3s/(s+3)  & \quad \text{if } s \le 3/2\\
    1       & \quad \text{if } s \ge 3/2 \,.\\
  \end{cases}
}

The results confirm our expectations on the
changes to the accretion shock radius (Fig.\;\ref{fig:lambd_sh} and
Table.\;3), and also shows that even modifying the inner
total mass profile from the self-similar solution to an NFW profile only
slightly increases the shock radius. For a first-order estimation,
the effect of relaxation and a moderate nonzero baryon content on
the accretion shock position can be neglected\footnote{On the other hand, the
effect of inner mass profile on the splashback radius is larger at high mass
accretion rates, as suggested by the dependence of the $r_{\rm sp}$ value on
the concentration parameter of NFW-shape halos \citep{shi16}.}.

\section{ICM properties}
\label{sec:ICMprof}
Analytical models based on a smooth accretion picture have been successful in
explaining some basic observed properties of the ICM, in particular its entropy profile
\citep{tozzi01,voit03}, despite the cluster growth being partly
clumpy rather than smooth. Here we show that, even with the
self-similar spherical collapse model which perhaps can be considered as the simplest smooth accretion model, some of the
observed ICM properties are already captured, and insights on their underlying physics can be obtained.

Since the gas is considered as adiabatic after the accretion shock in the
self-similar spherical collapse model, the effects of radiative cooling and feedback processes are
absent in our results. This needs to be kept in mind when interpreting the ICM
properties especially in the cluster core region where these effects are
significant.

\subsection[]{Gas mass profile}
The $M_{\rm gas}$ profile in the self-similar solution is close to a power-law
inside the accretion shock (Fig.\;\ref{fig:gasprofs_selfsim}). This guarantees
that the gas density, entropy, and pressure profiles are also approximately of
power-law shape (see the next subsection). 
For a small accretion rate $s\le 3/2$, the power-law slope $\alpha_{\rm M}$ of
$M_{\rm gas}$ is almost identical to that of $M_{\rm dm}$, because the
both the trajectory of a shell of gas and the apoapsis of the dark matter
trajectory approach a finite Eulerian radius (Fig.\;\ref{fig:traj2}).
For a larger accretion rate, the trajectory of a gas shell still stops at a
finite radius, but the orbit of a dark matter shell continues to shrink in
amplitude, and thus leads to a shallower mass slope than that of the gas.

\begin{figure}
\centering
    \includegraphics[width=.37\textwidth]{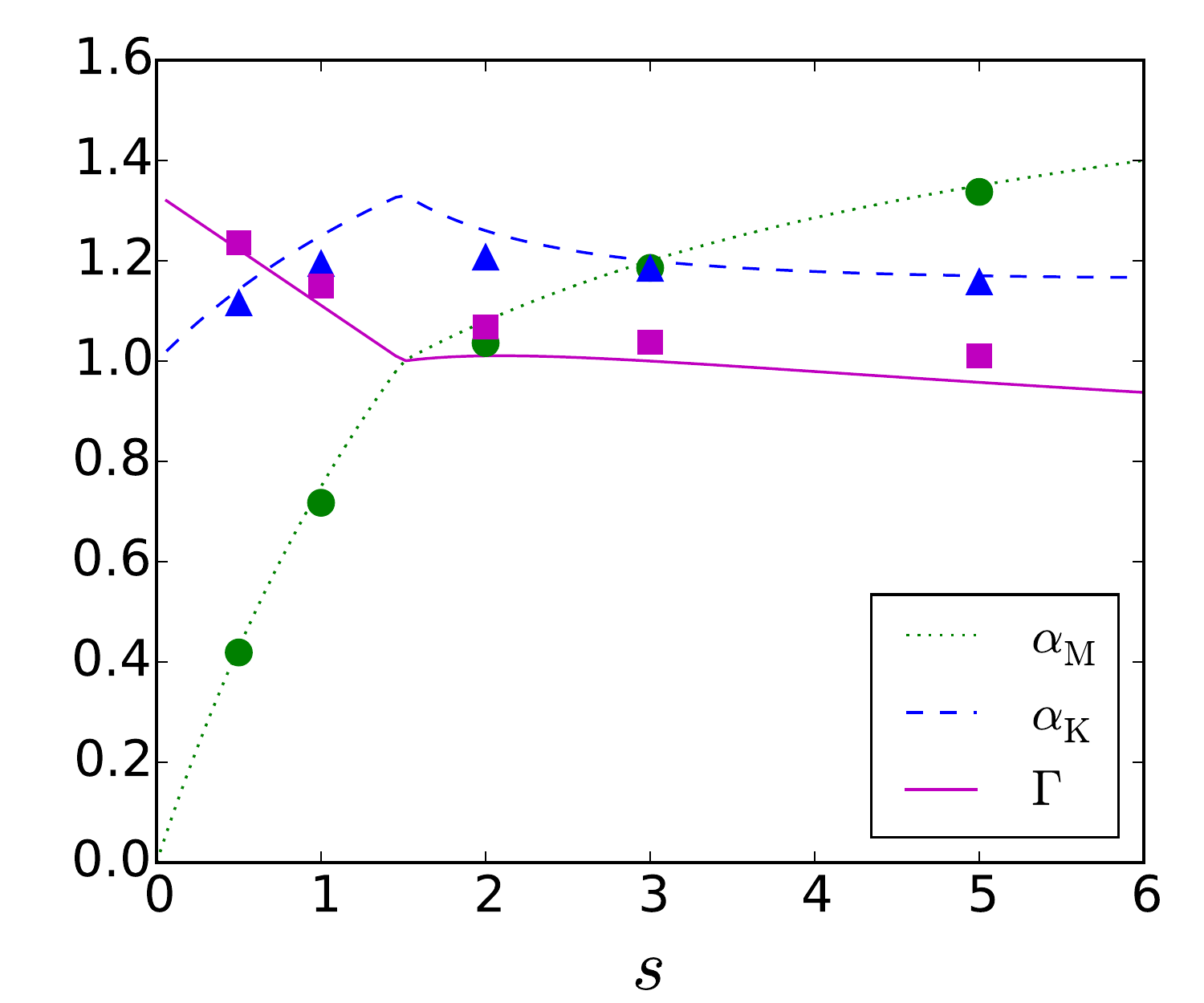} 
  \caption{Logarithmic slope for the gas mass profile (green), entropy profile
  (blue), and the effective polytropic index $\Gamma$ (magenta) as functions
  of the accretion rate $s$.
  The markers show values fitted using the self-similar solution on the
  radial range $0.01 \lambda_{\rm sh} < \lambda < 0.5 \lambda_{\rm sh}$. The
  lines show the analytical estimations given by Eqs.\;(\ref{eq:slope_M}), (\ref{eq:slope_K}) and
  (\ref{eq:slope_Gamma}).}
\label{fig:slopes}
\end{figure}

The above physical picture motivates us to find an approximation for
$\alpha_{\rm M}$ based on the asymptotic mass slope for the dark matter
$\Upsilon$, which takes account of the additional steepening of gas mass slope
compared to that of the dark matter at large accretion rates, 
\eq{
 \label{eq:slope_M}
  \alpha_{\rm M} \approx
  \begin{cases}
    3s/(s+3)  & \quad \text{if } s \le 3/2\\
    9(s+1)/(s+3)/5  & \quad \text{if } s \ge 3/2 \,.\\
  \end{cases}
}
The slope $\alpha_{\rm M}$ monotonously steepens with accretion rate $s$, and
ranges from 0.8 for $s=0.5$ to 1.4 for $s=5$ (Fig.\;\ref{fig:slopes}). For a
mass accretion rate $s > 3/2$, the mass slope depends also on the adiabatic
index of the gas, and the expression above is valid only for $\gamma=5/3$ (see
Appendix.\;\ref{app:ICM_gamma}).

\subsection[]{Entropy profile}
Being constant in adiabatic processes, entropy records the heating and cooling
history of the ICM. Thus the entropy profile is a very useful physical property that
describes the structure of the ICM. 

The key idea of the smooth accretion models is that the
entropy distribution in the ICM is primarily set by the cluster's mass accretion
history, i.e. the entropy and the mass of the gas are correlated.
Since the post-shock flow is adiabatic, the entropy of the gas is set at the accretion shock in a smooth
accretion model. The intracluster gas entropy can thus be used as a Lagrangian variable like the gas
mass $M_{\rm gas}$, and therefore a simple relation between the two
is expected. 

Combining the mass and entropy integrals (Appendices.\;\ref{sec:massint} and
\ref{sec:entint}) we find this relation to be
\citep[cf.][]{bert83} \eq{
\label{eq:KM}
K \propto M_{\rm gas}^{-\zeta}
}
in the self-similar model, with $-\zeta = {2(\gamma + \delta
-2)}/\br{3\delta-2} = (2s+3)/(s+3)$ varying from 2.7 for $s=0.5$ to 0.9 for
$s=5$. Note that this relation is derived simply from the continuity and
the adiabatic equations, without making use of the shock jump
conditions which is the basis of the $K(M, \dot{M})$ relation in the other
smooth accretion models \citep{tozzi01, voit03}.

Considering that the mass accretion rate may change (even significantly)
throughout the assembly of a galaxy cluster, a more generalized version of this
relation is  
$\dd \ln K / \dd \ln r = -\zeta (\dd \ln M_{\rm gas} / \dd \ln r)$, that the
local logarithmic slopes of the entropy and gas mass profiles are linearly
related with a pre-factor depending on the accretion rate. Since both the
entropy and the gas mass can be inferred directly from X-ray and millimetre
observations of galaxy clusters, it would be interesting to test this generalised relation in observations and
numerical simulations to see to which degree it holds in a more realistic
mass accretion scenario as well as when secondary effects such as cooling and
additional ways of entropy injection e.g. AGN feedback are present. In case the
relation holds well in more realistic conditions, testing how the relation
between entropy and gas mass slopes depends on the mass accretion rate in simulations may potentially lead to a method of inferring the mass accretion history.

With Eqs.\;(\ref{eq:slope_M}) and (\ref{eq:KM}) we can estimate the logarithmic
slope of the entropy profile as 
\eq{
\label{eq:slope_K}
\alpha_{\rm K} = -\zeta \alpha_{\rm M}\,. 
} 
Remarkably, although both $\zeta$ and the gas mass slope $\alpha_{\rm M}$ vary
quite significantly with the accretion rate, the value of the entropy slope
stays within a small range $\alpha_{\rm K} \sim 1.1 - 1.2$
(Fig.\;\ref{fig:slopes}). 
The inferred values of the entropy slope also match with that of the
`baseline intracluster entropy profile' \citep{voit05}, which presents the
theoretical expectation from a smooth accretion model as well as from adiabatic
hydrodynamical simulations. Many observed entropy profiles do follow a power-law
shape with a similar slope between the cluster core regions where they are
significantly modified by cooling and feedback processes, and the cluster
outskirts where the measurement becomes much harder and still controversial
\citep[e.g.][and referenced therein]{walker12,eckert13}.

\subsection[]{Effective equation of state}
Analytical models of the ICM profiles often make the approximation that the
intracluster gas has an effective equation of state $p \propto \rho^{\Gamma}$
where the value of $\Gamma$ does not depend on radius \citep{lea75, kom01,
shaw10}.
This is found by hydrodynamical numerical simulations to be precise to
$\sim$10\% in the virial regions of galaxy clusters
\citep[e.g.][]{shaw10,bat12b}, and the value for this effective polytropic index
$\Gamma \sim 1.1-1.2$ matches the observations \citep[e.g.][]{eckert13}.

For the self-similar solutions, the approximate effective polytropic index can
be estimated given $\alpha_{\rm K}$ and the logarithmic slope
of the density profile $\alpha_{\rm D} = \alpha_{\rm M} -3$, as 
\eq{
\label{eq:slope_Gamma}
\Gamma =
\alpha_{\rm K}/\alpha_{\rm D} + \gamma = \gamma -\zeta \alpha_{\rm M} /
(\alpha_{\rm M} -3)\,.
} 
Like $\alpha_{\rm K}$, the approximate values for
$\Gamma$ also stay within a small range when the accretion rate varies
(Fig.\;\ref{fig:slopes}). In detail, $\Gamma$ slightly decreases at larger
accretion rates. This is in agreement with the weak positive correlation
between $\Gamma$ and the concentration parameter of the dark matter halo found
in \citet{kom01,kom02}\footnote{In fact, using the method described in
Sect.\;\ref{sec:diffprof} to match the accretion rate and the concentration
parameter, our accretion rate dependence of $\Gamma$ agrees with the
\citet{kom02} fitting formula (their equation 17) both in their
trends and also quantitatively to $<$5\% 
for $0.5\le s \le 5$. Although we cannot rule out the possibility of this being
a coincidence, this makes it more interesting to look for this weak but
monotonous dependence of $\Gamma$ on the accretion rate or on the concentration
parameter in hydrodynamical simulations and observations.}.

This effective polytropic index $\Gamma$ differs physically from the polytropic
index, or the adiabatic index of the gas $\gamma$. While the latter reflects
how the gas pressure changes under compression or expansion, the former is
merely an effective description of the global ICM structure. Nevertheless, one
may wonder why the ICM structure is characterised by an effective polytropic
index $\Gamma < \gamma$. From the relation Eq.\;(\ref{eq:slope_Gamma}) above,
the value of $\Gamma$ is set by the $K(M_{\rm gas})$ relation which describes
the entropy generation during mass accretion, and the gas mass profile
itself\footnote{This is also noticed in
\citet{voit03}. However, a $\Gamma$ value of $\sim 1.1-1.2$ does not require a
near linear relation between $K$ and $M_{\rm gas}$ as stated there.}.
That $\Gamma < \gamma$ is simply due to a positive slope of the entropy profile
$\alpha_{\rm K}$ as a result of an increasing entropy production at the
accretion shock with growing mass of the cluster.

\subsection[]{Gas mass fraction}

\begin{figure}
\centering
    \includegraphics[width=.37\textwidth]{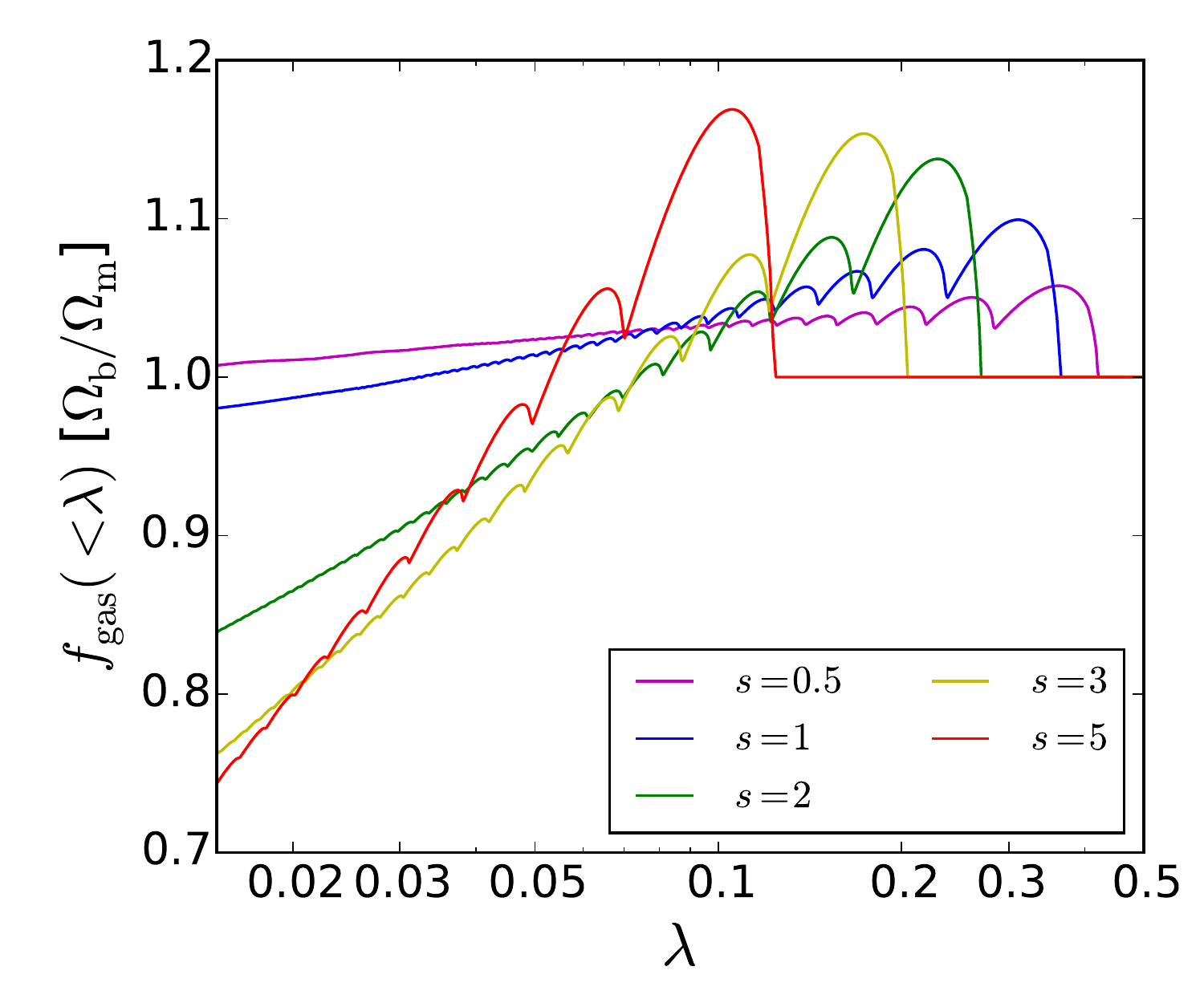} 
  \caption{Gas mass fraction of the self-similar solutions for various accretion
  rates (lines with different colors). The values are normalised to the cosmic
  mean gas mass fraction $\Omega_{\rm b}/\Omega_{\rm m}$.}
\label{fig:fgas}
\end{figure}

In observations, the gas mass fraction $f_{\rm gas} = M_{\rm gas}/M$ is
typically found to be less than the cosmic mean value at small radii, and approaches the cosmic mean around
the virial radius of the cluster \citep[e.g.][]{pratt10,eckert13}. This has been
commonly regarded as evidence for feedback by e.g. AGNs
in the central galaxies which expels the inner baryonic material to large radii and even out of the potential well of the halo. 
Such an interpretation is supported by the observation of progressively more
significant baryon loss from clusters of galaxies, groups of galaxies to galaxies themselves
\citep[e.g.][and references therein]{dai10}.

Here we consider an additional cause to the deviation of the gas mass
fraction from the cosmic mean value: that gas lagging behind dark matter during
its infall on to the cluster after going through the accretion shock.
Fig.\;\ref{fig:fgas} shows the gas mass fraction in the self-similar solutions
as influenced by the accretion shock. Unlike the entropy
profile and the effective equation of state of the ICM, we find the gas mass
fraction to depend strongly on the mass accretion rate.
The higher the mass accretion rate, the stronger the depletion of gas
relative to the dark matter in the inner regions of the cluster. The
depletion is already more than 20\% at $0.1 r_{\rm sh}$ for a mass accretion
rate $s \ge 3$. This is again caused by the stronger adiabatic contraction
at high accretion rates, which steepens the gravitational potential in the
central regions and causes the orbits of the dark matter to contract further in
comparison with the gas. Since the effect of this accretion dynamics on the
$f_{\rm gas}$ is most likely secondary to that of the AGN feedback, the $f_{\rm
gas}$ profiles presented here may not be realistic. Nevertheless, the strong
dependence of the inner gas fraction on the mass accretion rate found here may
be observable, which encourages one to study this dependence in numerical
simulations and in observations.

\section[]{Discussions}
\label{sec:discussion}
\subsection[]{Effect of deviation from smooth accretion}
In the hierarchical structure formation picture, galaxy clusters grow not all
via smooth accretion as assumed in this paper and by other smooth accretion
models. Mergers and accretion through filaments are also significant
modes of accretion. They destroy the Lagrangian nature of the radial coordinate
by mixing matter accreted at different times, and they introduce density
inhomogeneities to the ICM. Quantitative effects of these are hard to
evaluate and remain an open question.

\citet{voit03} addressed the effect of density inhomogeneities in the material
being accreted on the entropy of the ICM. They find that a homogeneous accretion
maximizes the observed entropy as a pure effect of mass-weighting during the
averaging.
Similarly, mass-weighted accretion shock radius would be biased towards that
of the higher density gas. As gas with slightly higher density carries more
momentum per volume, it would get shocked at a smaller radius, and thus density
inhomogeneities would lead to a smaller mass-weighted accretion shock radius.

Mixing of matter accreted at different times tends to wash out the dependence of
the ICM profiles on the mass accretion history. Supposedly, it would at the same time drive the profiles
towards a more `relaxed' state. However, since the ICM structure given by smooth
accretion models with a typical mass accretion history is generally already
stable and smooth, the effect of this radial mixing may not be prominent in
reality.

\subsection[]{Effect of Dark Energy}
\label{sec:DE}
Throughout the paper we have been studying the accretion shock location and the
ICM profiles in an Einstein de-Sitter universe where self-similarity strictly
holds. How would our results change in a $\Lambda$CDM universe with a non-zero
dark energy content? 

The effect of dark energy is mainly on the dynamics of dark matter and gas
before they turn around. Within the turn-around radius, the gravity of the halo
dominates and the role of dark energy is negligible. In this sense, the effect
of dark energy on the properties of the cluster region is through the initial
condition at turn-around, and can partly be mimicked by a change of the mass
accretion rate $s$. Thus, we expect the features that have little dependence on $s$ to
remain in a $\Lambda$CDM universe, such as that the logarithmic slope of entropy
profile and the effective polytropic index of the intracluster gas lying in a
narrow range of ~1.1-1.2, and that the shock radius aligns well with the dark
matter splashback radius for moderate accretion rates $s \lesssim 5$.
The latter, combined with the origin of the good scaling between splashback
radius and $r_{\rm 200m}$ \citep{shi16}, would also suggest that the accretion shock
scales well with $r_{\rm 200m}$ for a similar reason: the
correlated increase of the matter content of the universe and the average halo
mass accretion rate with redshift, and their canceling effects on $r_{\rm sh} /
r_{\rm 200m}$.

Quantitative estimations of the effect of dark energy requires an extension of
the current framework to a $\Lambda$CDM universe, where the evolution of
the cosmological background introduces an additional time scale to the system
and breaks the strict self-similarity. In this case, we can consider the change
of the cosmological background as described by e.g. the matter
content $\Omega_{\rm m}$ as an adiabatic variable, and approximate the
relatively fast dynamics of gas and dark matter as self-similar for each $\Omega_{\rm m}$ value. The extension for the
dark matter has already been achieved in \citet{shi16} and have yielded results
confirming our expectations. We leave the extension for the gas to future work.

\section[]{Conclusion}
\label{sec:con}
Although simple and fully analytical, the self-similar spherical collapse model
already captures some fundamental ICM properties such as the location
of the accretion shock, its alignment with the splashback radius, the entropy
profile and the effective polytropic index of the intracluster gas. 
We have systematically studied the ICM properties given by the self-similar
spherical collapse model for dark matter and gas by extending 
Bertschinger's work to various mass accretion rates. We also consider the
effect of a different inner mass profile due to dynamical relaxation, and the effect of a
moderate baryon content. 

Our main findings can be summarised as follows. 
\items{
\item  The radial position of the accretion shock is set by the continuities of
mass, momentum and enthalpy between the approximately hydrostatic cluster gas
and the gas being accreted by the cluster. Neither a modification of the inner
mass profile to an NFW shape due to dynamical relaxation nor the exact value of
the baryon content has much effect on the accretion shock location.

\item The alignment of the accretion shock
radius and splashback radius discovered in hydrodynamical simulations is
\textit{not} universal, in the sense that it holds only for a gas with an
adiabatic index of $\gamma \approx 5/3$ and for not-too-high mass accretion
rates. For the intergalactic gas for which $\gamma \approx 5/3$, the alignment
indeed holds for a large range of low and moderate accretion rates $s \lesssim
5$ which are typical for galaxy clusters at low redshifts. This arises from the
dependencies of the accretion shock and the splashback radius on the accretion
rate, which are the same in direction and similar in degree, but different in
physics. 
For the gas, it is because the higher energy and momentum of
the inflowing gas associated with a higher mass accretion rate. For the
dark matter, it is due to a more significant halo growth during the time between the splashback and when the matter at
splashback was accreted. 

\item In the self-similar spherical collapse model, like in any smooth accretion
model, the entropy distribution in the ICM is set by the cluster's
mass accretion history. However, the logarithmic slope of the entropy profile is
rather independent of the mass accretion rate and lies around $1.1-1.2$. This
confirms the robustness of the `baseline intracluster entropy profile'
of \citet{voit05}. 

\item A linear relation exists between the local logarithmic slopes
of the entropy and gas mass profiles, with a pre-factor depending on the mass
accretion rate. Although based on a heavily simplified picture, this relation is
worth testing in hydrodynamical simulations and observations, and may lead to a
method of constraining the mass accretion history of galaxy clusters from
observations of the ICM.

\item  This effective polytropic index of the intracluster gas is
an effective description of the global ICM structure, and is set by the gas mass
profile and the entropy generation during mass accretion. It also has a
value around $1.1-1.2$ and is rather independent of the mass accretion rate. Its
value is smaller than that of the actual polytropic index (the adiabatic index)
of the gas as a result of the cluster mass growth and the consequent increasing
entropy production at the accretion shock.

\item The accretion shock slows down the gas compared to the dark matter during
their accretion on to a galaxy cluster, and leads to a deficit of gas mass
fraction compared to the cosmic mean in the inner regions of the cluster. Unlike
the entropy profile and the effective polytropic index of the ICM, this effect
depends strongly on the mass accretion rate of the cluster. This calls for
simulation and observation studies of $f_{\rm gas}$ as a function of the mass accretion rate.
}

\section*{Acknowledgements}
XS is grateful to Eiichiro Komatsu for carefully reading the manuscript and
giving helpful suggestions, as well as to Erwin Lau for discussions and
comments, and to Daisuke Nagai for related discussions during his visit at MPA.

\bibliographystyle{mn2e}

\appendix

\section{Integrals of motion}

\citet{bert85} has identified three integrals of motion for the self-similar
spherical collapse solution of dark matter and gas with $\Omega_{\rm
b}\ll 1$. Apart from providing checks for the numerical solution, we find them
valuable for understanding the shapes of the ICM profiles
(Sect.\;\ref{sec:ICMprof}).
Here we derive these integrals of motion for an arbitrary accretion rate $s$.

\subsection{Mass integral}
\label{sec:massint}
To keep self-similarity, the fraction of gas mass within a radius $\lambda$
should stay constant with time. This gives a relation between the gas mass and
its flux \citep[cf.][]{bert83}, 
\eq{
\label{eq:mflux}
\frac{m_{\rm gas}}{m_{\rm gas, ta}} = \frac{4\pi r^2\rho (\lambda v_{\rm ta} -
v)}{4\pi r_{\rm ta}^2 \rho_{\rm ta} v_{\rm ta}} \,.
}
Using Eq.\;(\ref{eq:nondim}), and that 
\eq{
\rho_{\rm ta} = \frac{m_{\rm gas, ta}}{4\pi r_{\rm ta}^3} \frac{\dd \ln m_{\rm
ta}}{\dd \ln t} \br{\frac{\dd \ln r_{\rm ta}}{\dd \ln t}}^{-1} =
\frac{m_{\rm gas, ta}}{4\pi r_{\rm ta}^3} \frac{2s}{3\delta}
}
to express Eq.\;(\ref{eq:mflux}) in terms of the nondimensional
quantities, it is \eq{
\label{eq:intconst_M}
M_{\rm gas} = - \frac{9}{2s} \lambda^2 D \bar{V} \,,
}
with $\bar{V} = V - \delta \lambda_{\rm sh}$ being the velocity in the frame
where the shock is static.

\subsection{Entropy integral}
\label{sec:entint}
We re-write the continuity and adiabatic equations of the gas (the first and the
third equations of the fluid equations Eq.\;\ref{eq:fluid}) as
\eqs{
\label{eq:nondim_ln}
& (\ln \lambda^2 D \bar{V})'  =  - \frac{3\delta-2}{\bar{V}} \,,
\\
& (\ln K)' = (\ln P)' - \gamma (\ln D)' = - \frac{{2(\gamma+\delta-2)}}{\bar{V}}
\,. 
}
In this form, it is straightforward to derive that 
$K \br{\lambda^2 D\bar{V}}^{\zeta}=$const, 
with
\eq{
\zeta = -\frac{2(\gamma + \delta -2)}{3\delta-2} = -\frac{2}{3} - \frac{1}{s}
\,. }

\subsection{Virial theorem}
\begin{figure}
\centering
    \includegraphics[width=.47\textwidth]{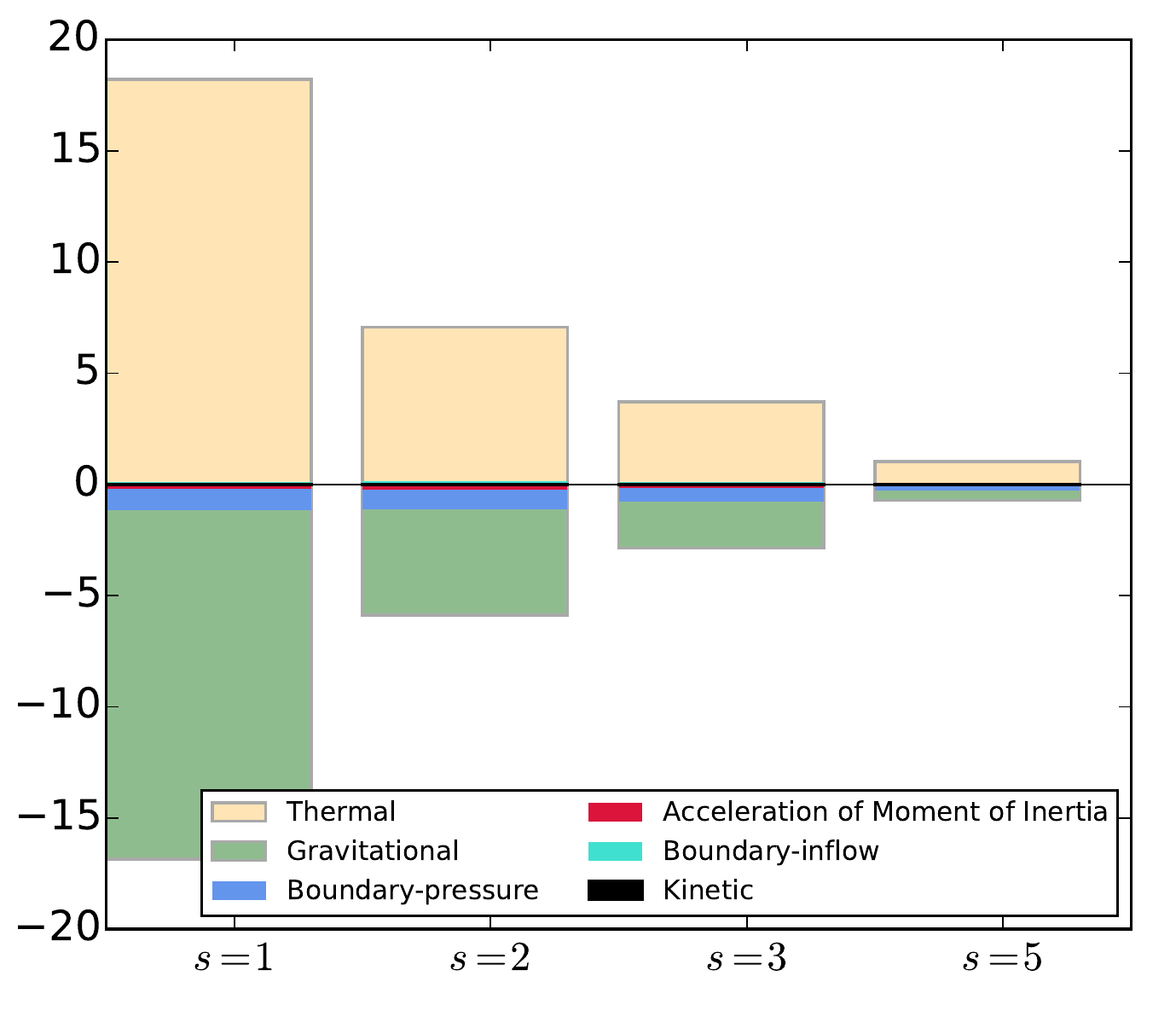} 
  \caption{Test of the virial theorem in the shock-enclosed region. Kinetic,
  thermal, gravitational energy terms of Eq.\;(\ref{eq:virial}) are shown,
  as well as the acceleration of the moment of inertia, and the boundary terms
  from mass inflow and gas pressure. The $s=0.5$ case is not shown because
  there the thermal and gravitational energies diverge at $r=0$.}
\label{fig:virial}
\end{figure}

The two integrals of motion above are local ones which is valid at each
radius. Another global integral of motion is provided by the virial theorem. 
The dimensionless specific kinetic, thermal and gravitational energies averaged
within the accretion shock are \citep{bert85}
\eqs{
& t = \int_0^{M_2} \frac{V^2}{2} \dd M \\
& u = \int_0^{M_2} \frac{P/D}{\gamma-1} \dd M \\
& w = -\frac{2}{9} \int_0^{M_2} \frac{M_x}{\lambda} \dd M  \,. 
}
The virial theorem can be derived from the Euler equation in
Eq.\;(\ref{eq:fluid}) as 
\eqs{
\label{eq:virial}
& \br{\frac{25}{2}\delta^2 - \frac{25}{2}\delta + 3} \int \lambda^2 \dd M \\
=& 2t
+ 3(\gamma-1)u +w  \\
& - 3 \lambda_s^3 D_2 \bar{V}_2 \bb{\bar{V}_2 +
\frac{7\delta-3}{2}\lambda_s}  - \frac{6}{\gamma-1} \lambda_s^3 D_2\bar{V}_2^2
\,.}
The l.h.s. is $\dd^2 I /\dd t^2 /2$ where $I$ is the moment of inertia. The two
boundary terms on the r.h.s. can be interpreted as originating from the flux of
inertia through the boundary and the pressure at the boundary respectively
\footnote{The second boundary term is missing in the virial theorem expression
in \citet{bert85}. It is larger than the first boundary term in magnitude
(Fig.\;\ref{fig:virial}).}.
The second boundary term $- {6} \lambda_s^3 D_2\bar{V}_2^2 / \br{\gamma-1}$ can
be also expressed as $-3 P_2 \lambda_s^3$.

With the numerical self-similar solution presented in this paper, the sum of all
terms is approximately zero with a deviation less than one percent of the
thermal energy. As shown in Fig.\;\ref{fig:virial},
the thermal and gravitational energy are the dominating terms in the virial
theorem. The other terms are (in decreasing order of their magnitudes): the
second boundary term, the acceleration of the moment of
inertia, the first boundary term, and the kinetic energy.

\section{Influence of the adiabatic index on the ICM profiles}
\label{app:ICM_gamma}

\begin{figure*}
\centering
    \includegraphics[width=.97\textwidth]{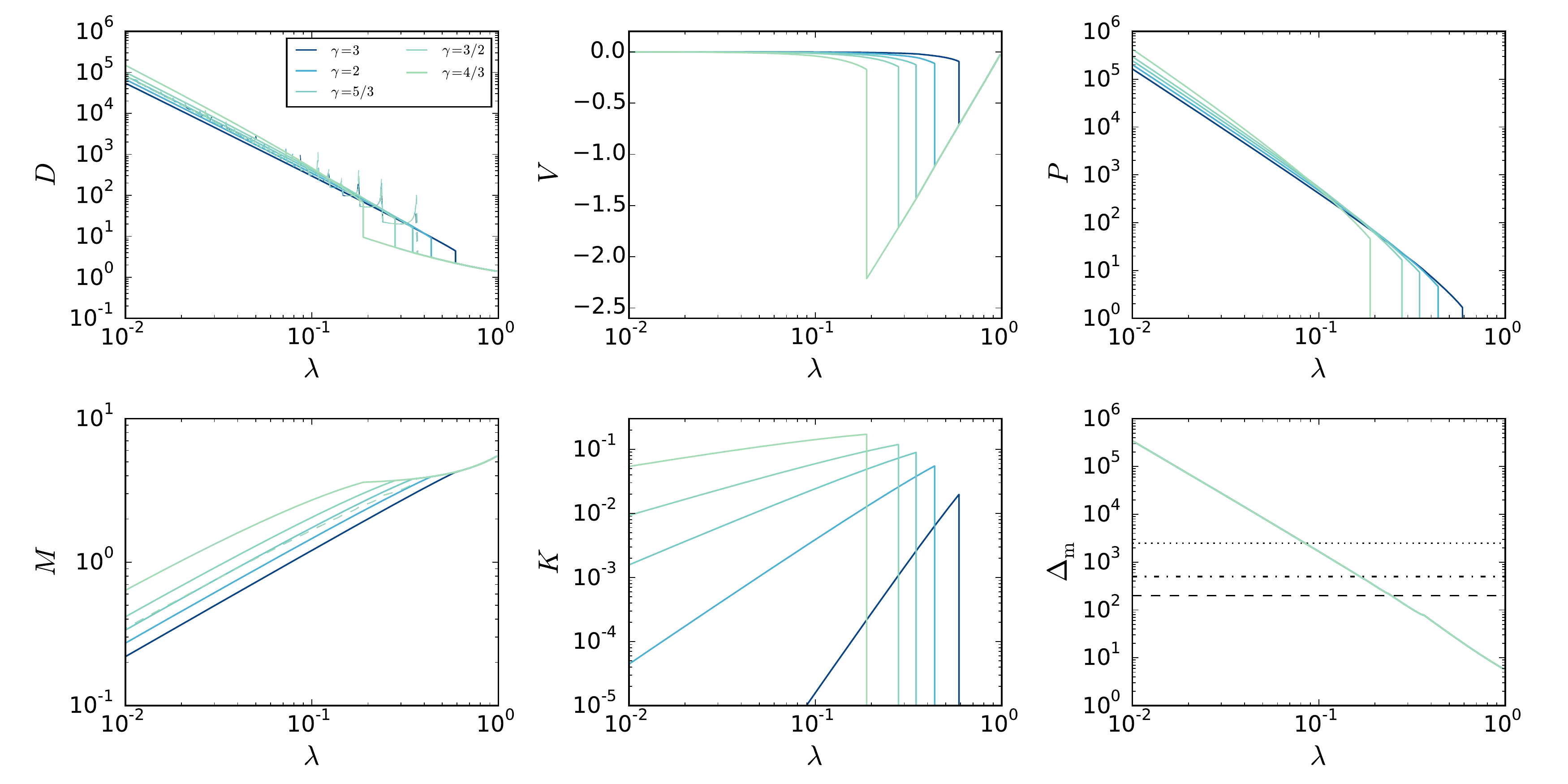} 
  \caption{Dimensionless ICM profiles of the self-similar spherical collapse
  model. Similar to Fig.\;\ref{fig:gasprofs_selfsim} but showing the dependency
  of the profiles on the adiabatic index $\gamma$ of the gas. A mass accretion
  rate of $s=1$ is used.}
\label{fig:gasprof_gamma}
\end{figure*}

\begin{figure*}
\centering
    \includegraphics[width=.97\textwidth]{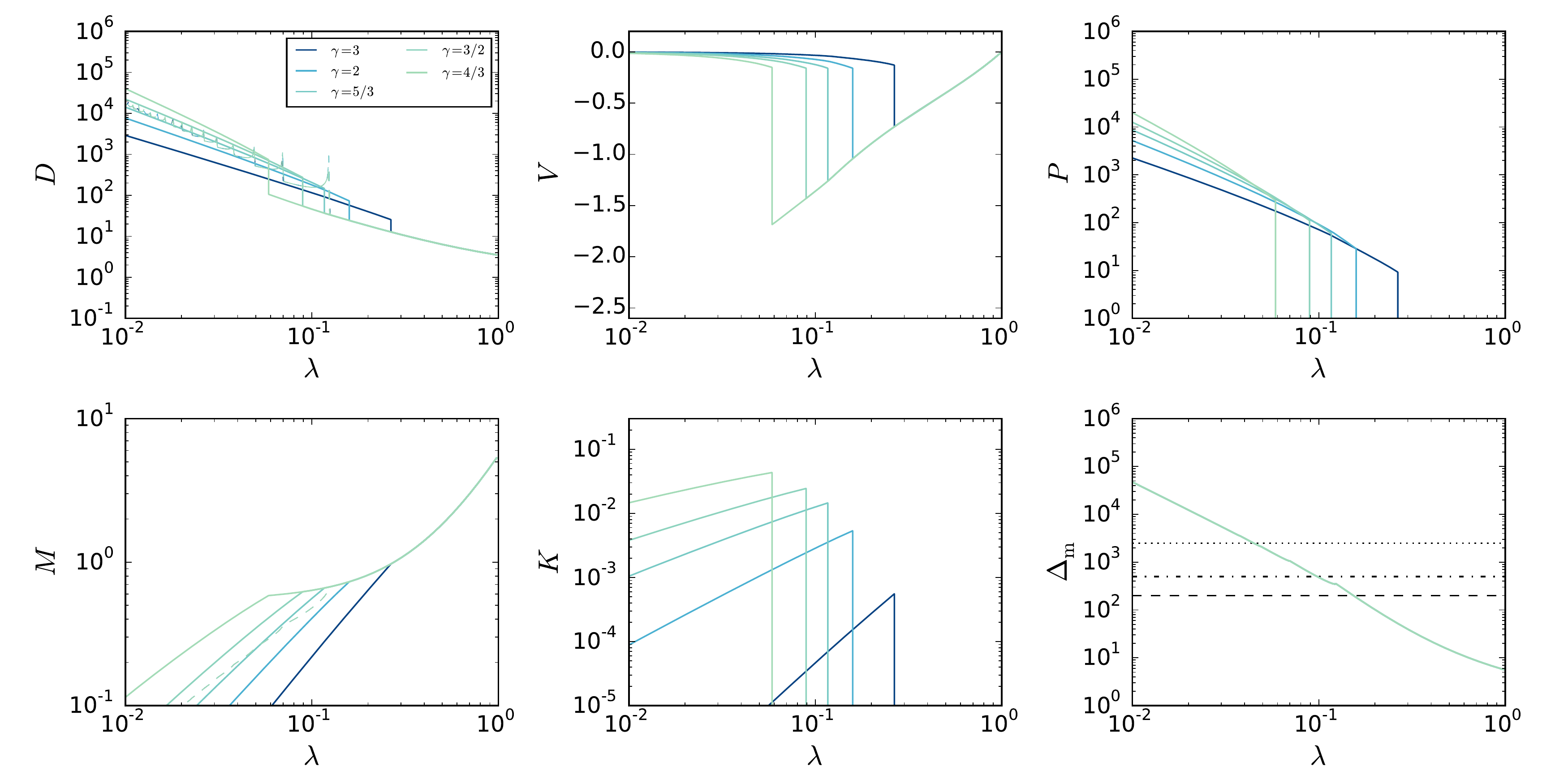} 
  \caption{Same as Fig.\;\ref{fig:gasprof_gamma} but for a mass accretion rate
  of $s=5$.}
\label{fig:gasprof_gamma5}
\end{figure*}

The diffuse intergalactic gas that accretes on to galaxy clusters is
composed dominantly of ionized single-atom particles, and thus has an adiabatic
index $\gamma=5/3$. Nevertheless, exploring how the ICM properties depend on the
adiabatic index can lead to a better understanding of the physical origin of the
ICM profiles. Also, as a means of entropy reduction, the effect of radiative
cooling can be mimicked by a softer (smaller) adiabatic index; as an additional
entropy injection mechanism, AGN feedback can be mimicked by a stiffer
(larger) adiabatic index. Thus studying cases for gas with
different adiabatic indices also allows one to see roughly what effects cooling
and AGN feedback have on the ICM profiles.

Figs.\;\ref{fig:gasprof_gamma} and \ref{fig:gasprof_gamma5} present the
self-similar ICM profiles for gas with various adiabatic indices ranging from
$\gamma=4/3$ that corresponds to a gas with six degrees of freedom e.g. a
diatomic gas at high temperature, to $\gamma=3$ that corresponds to a gas
with only one degree of freedom. Except for the change on the accretion shock
location that has been discussed in Sect.\;\ref{sec:rshock}, the most
prominent effect of $\gamma$ on the ICM is a steeper entropy profile at a stiffer adiabatic index,
which is reflected in the adiabatic equation e.g. the second equation in
Eq.\;(\ref{eq:nondim_ln}). Another influence of different $\gamma$ values is
prominent only at high accretion rate: the gas mass profile also steepens with a
stiffer adiabatic index when $s>3/2$. The gas density and pressure profile
slopes, the effective polytropic index, and the gas mass fraction are also
affected as a consequence (see Sect.\;\ref{sec:ICMprof}).

\end{document}